# Asymmetric magnetization splitting in diamond domain structure: Dependence on exchange interaction and anisotropy

Kaixuan Xie,[1] Xiaopu Zhang,[1] Weiwei Lin,[2*] Peng Zhang,[1] and Hai Sang[1†]

[1]*National Laboratory of Solid State Microstructures, School of Physics, Nanjing University, Nanjing 210093, China*

[2]*Institut d'Electronique Fondamentale, Université Paris-Sud, Orsay 91405, France*

The distributions of magnetization orientation for both Landau and diamond domain structures in nano-rectangles have been investigated by micromagnetic simulation with various exchange coefficient and anisotropy constant. Both symmetric and asymmetric magnetization splitting are found in diamond domain structure, as well as only symmetric magnetization splitting in Landau structure. In the Landau structure, the splitting angle increases with the exchange coefficient but decreases slightly with the anisotropy constant, suggesting that the exchange interaction mainly contributes to the magnetization splitting in Landau structure. However in the diamond structure, the splitting angle increases with the anisotropy constant but derceases with the exchange coefficient, indicating that the magnetization splitting in diamond structure is resulted from magnetic anisotropy.

## I. INTRODUCTION

Magnetic nanostructures is an active study aspect, not only for its high sensitivity function[1] or excitations of magnetic moments for practical application,[2,3] but also as good candidates in research on some fundamental and interesting magnetism physics.[4–8] On the other hand, the framework of micromagnetic simulation can describe the magnetization configuration and dynamics in a scale between several tens nm and a few microns.[4,5,9] Basically there are two ways for micromagnetic calculations. One is based on the integration of the equation, which is the motion of the magnetic moments described by the Landau-Lifshitz-Gilbert (LLG) equation. The other one is to minimize the total magnetic energy. The former is in favor of showing time-dependent evolution of magnetization,[10] while the latter is in a perspective of energy. It shows that the size,[11,12] the aspect ratio[13,14] and the thickness[11,14] are important parameters over magnetic configuration.[9] In order to reduce the exchange energy, a uniform magnetization is ideal, but the demangetiztion energy would like magnetization parallel to surface or poles counter balance at interface. Meanwhile, magnetic moment favors low anisotropy energy if its magnetization is parallel to the easy direction. If no external field is applied, the magnetization will relax to an equilibrium state, which is the competition and compromie around these three energy items and reach a local minimum energy. No matter spatial dimension differences[11–15] or roughness variety,[16,17] in a viewpoint of minimum energy, they all act on the local and non-local magnetic energy items.[18]

In thin soft elements, the Landau state and the diamond state are well known as typical magnetic flux-closure patterns and prominent candidate of ground state.[4,5] In Ref. 4, the magnetization orientation splitting has been observed in the flux-closure domain which means the majority of the magnetizaton orientation in splitting domain is slightly away from the axis within the rectangle plane, and its roughness dependence has been explained successfully by micromagnetic simulation. However, the magnetization splitting was only reported in the Landau structure and it is symmetric.[4] In this paper, we report that both symmetric and asymmetric magnetization splitting is investaged in diamond domain structure. From the dependences of the splitting angle on the exchange coefficient and the anisotropy constant, it suggests that the exchange interaction mainly contributes to the magnetization splitting in Landau structure, while that in diamond structure is dominated by magnetic anisotropy.

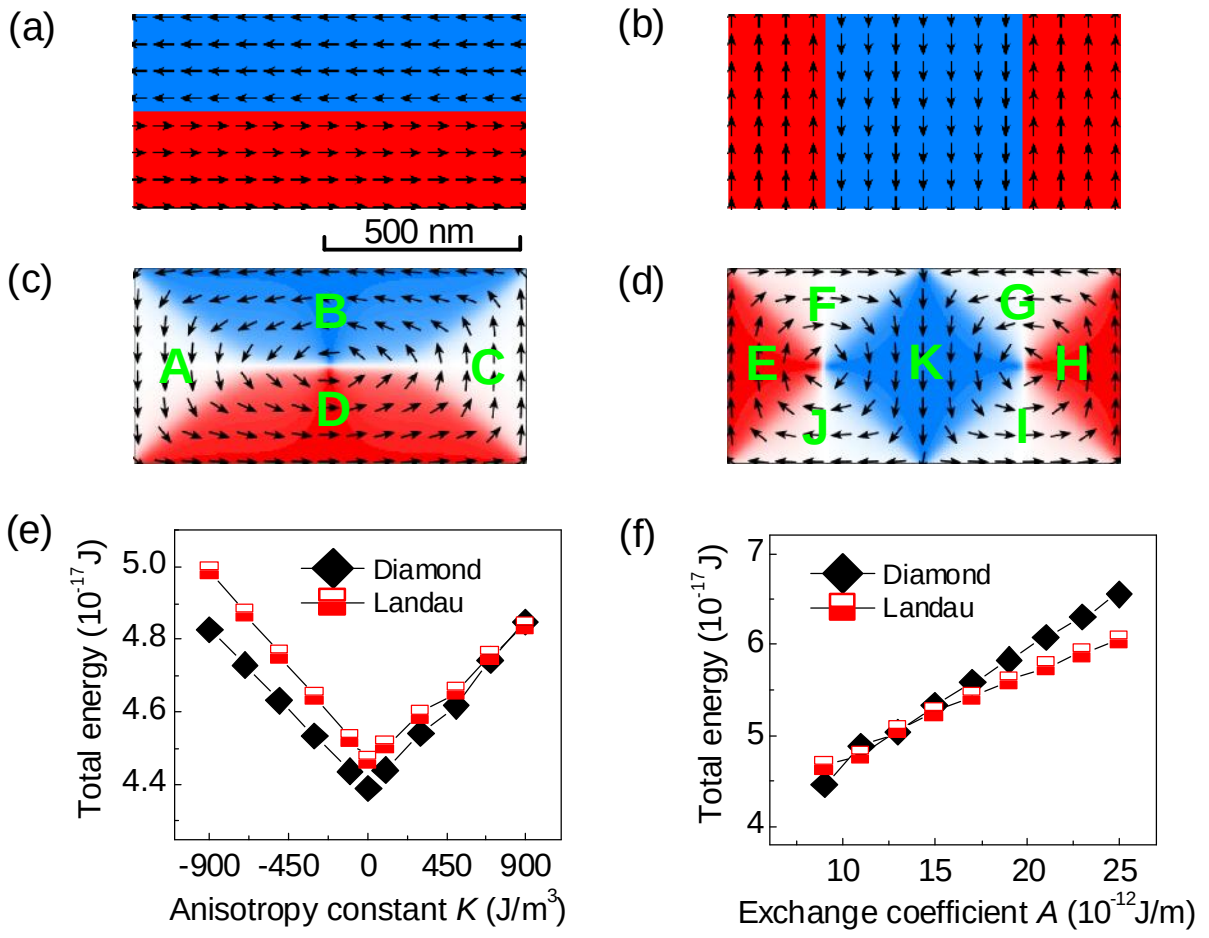

FIG. 1 (color online) (a) and (b), two initial magnetization states for relaxing to Landau domain structure (c) and diamond domain structure (d), respectively. Blue and red represent two magnetic domains, and the arrows show the magnetization orientation. (e) Dependence of total energy on anisotropy constant *K*. (f) Dependence of total energy on exchange coefficient *A*.

## II. MODEL

We made a systematical investigation on the magnetization orientation splitting in both Landau and diamond domain structures in nano-rectangles by OOMMF micromagnetic simulation, based on a model of Standard Problem 1.[19] The dependence of magentization splitting on exchange coefficient and anisotropy constant were studied. Principle of minimum energy and the symmetry of magnetic microstructure[16,20] were employed to explain the different dependence and distribution features. The simulated sample size is 1000 × 500 × 20 nm$^3$ with cell size no larger than 5 × 5 × 5 nm$^3$. In our simulation, no external magnetic field was applied. The total energy $E$ includes the exchange energy $E_{ex}$, anisotropy energy $E_A$ and demagnetizing energy $E_D$, where

$$E_{ex} = \int_{(V)} A[(\nabla m_x)^2 + (\nabla m_y)^2 + (\nabla m_z)^2]dV, \quad (1)$$

$$E_A = \int_{(V)} K(a_1{}^2 a_2{}^2 + a_1{}^2 a_3{}^2 + a_2{}^2 a_3{}^2)dV, \quad (2)$$

$$E_D = -\frac{m_0}{2}\int_{(V)} \boldsymbol{M}\ \boldsymbol{H}_d dV. \quad (3)$$

In the simulation, the minimization of total energy is performed by using the conjugate gradient method with no preconditioning,[21,22] by locating local minimum in the energy surface. The parameters of permalloy are used here as reference. The magnetization $M$ is 8.6 × 10$^5$ A/m, the exchange coefficient $A$ between cells varies from 9 × 10$^{12}$ J/m to 2.3 × 10$^{13}$ J/m, and the anisotropy constant $K$ between −900 J/m$^3$ and 900 J/m$^3$, respectively. Different magnetization configurations are chosen as the initial state without considering the thermal acitivation. Then, it

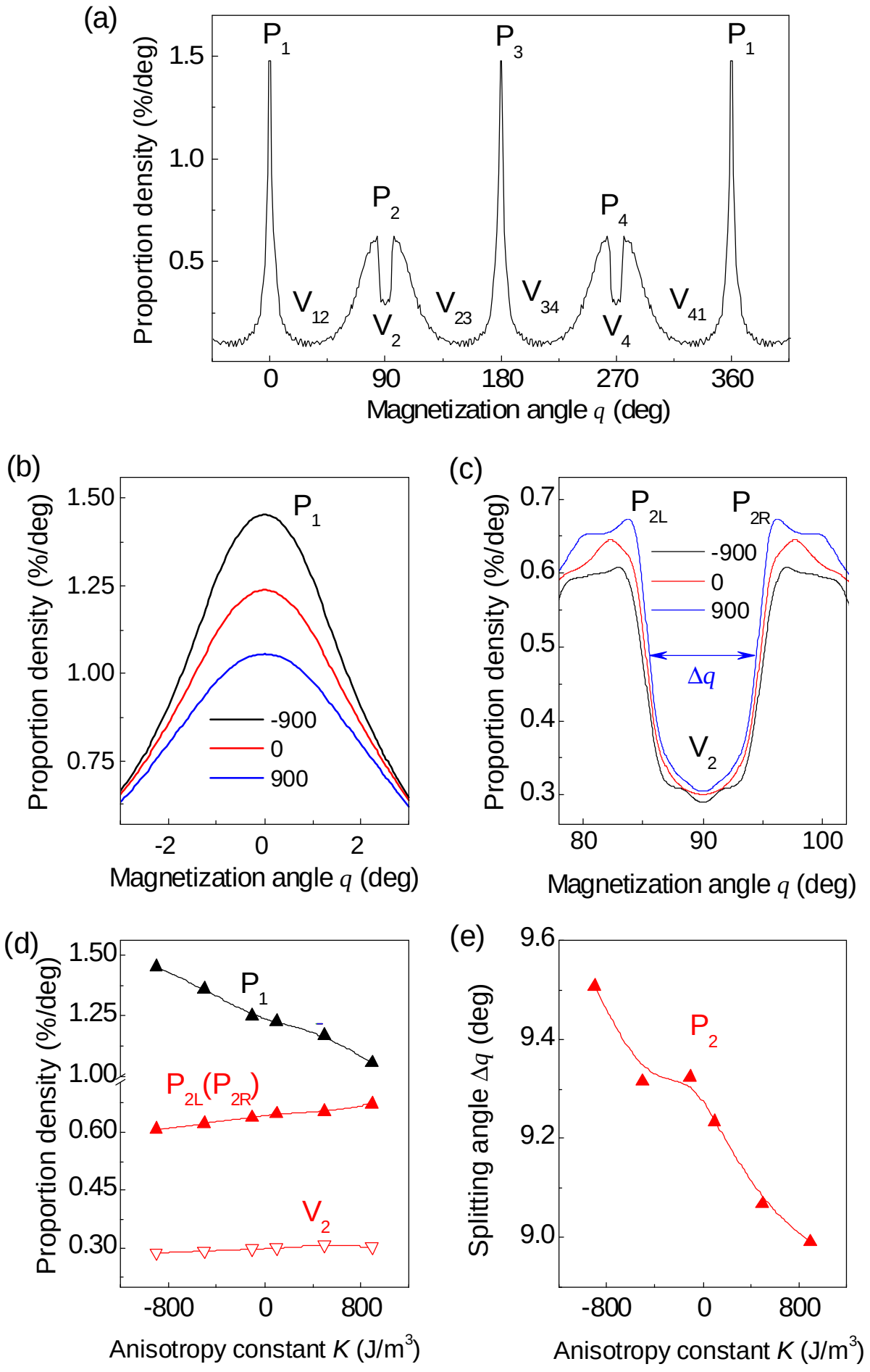

FIG. 2 (color online) For Landau domain structure: (a) the distribution curve of magnetization orientation as a function of the mangetization angle $\theta$. (b) the distribution curves of $P_1$ for $K$ = −900, 0 and 900 J/m$^3$, as $A$ = 1.3 × 10$^{13}$ J/m. (c) the distribution curves of $P_2$ for $K$ = −900, 0 and 900 J/m$^3$, as $A$ = 1.3 × 10$^{13}$ J/m. (d) Dependence of the proportion densities on $K$ for $P_1$, $P_2$ and $V_2$. (e) Dependence of the splitting angle $\Delta\theta$ on $K$ for $P_2$.

converges to a minimum energy state confined by the local energy barrier.[14] For an initial magnetization state described in Fig. 1(a), it relaxes to the Landau state as shown in Fig. 1(c); while a diamond state will form in the case of an initial state in Fig. 1(b), as shown in Fig. 1(d). The up direction of the short axis of the rectangle is defined as 0° direction. Positive anisotropy constant $K$ means the magnetizaton prefers in 90° or 270°, i.e. the long axis. Either the diamond structure or the Landau structure is converged to, although the total energy of each differs for different $A$ and $K$. The dependence of the total energy on $A$ and $K$ are shown in Fig. 1(e) and Fig. 1(f) respectively. It is found that mostly the lower total energy prefers the diamond structure as $K$ changes. This is similar to the results in the literatures.[12] However, the lower energy favors the Landau structure as $A$ larger than 1.5× 10$^{13}$ J/m.

## III. LANDAU DOMAIN STRUCTURE

Figure 2(a) shows a typical distribution curve of magnetization orientation as a function of the mangetization angle $\theta$ in Landau domain structure. The proportion density $P$ is calculated as the count in a step of 1.2° over the total cell amount. The distribution curve of magnetization orientation show a minimal period of 180°, corresponding to the centro-symmetry in Landau domain structure. Four main peaks $P_1$, $P_2$, $P_3$ and $P_4$ correspond to four different magnetization orientations in domain C, B, A and D respectively in Fig. 1(c). $P_2$ ($P_4$) splits to two symmtric peaks $P_{2L}$ and $P_{2R}$ ($P_{4L}$ and $P_{4R}$) with a valley $V_2$ ($V_4$), which correspond to the magnetization orientations in domain B (D) aligning with the long edges.[2] $V_{23}$ is the valley between $P_2$ and $P_3$, which locates in the 90° domain wall between domain A and B. Hereafter, we name all the peaks and valleys in this way. Figure 2(b) show the distribution curves of $P_1$ for $K$ = −900, 0 and 900 J/m$^3$, in the case of $A$ = 1.3 × 10$^{13}$ J/m, and those of $P_2$ are shown in Fig. 2(c). The splitting angle $\Delta\theta$ between $P_{2L}$ and $P_{2R}$ defined by the half depth of the valley is about 9.5° (see Fig. 2(c)). The proportion densities of $P_1$ decreases as $K$ increasing, while those of $P_{2L}$, $P_{2R}$ and $V_2$ increase with $K$, as shown in Fig. 2(d). It indicates that there is a net increment of magnetization in domain B but a decrement of that in domain C. More magnetization aligns close to the easy axis for benefiting a low total energy when the amplitude of anisotropy is larger. It can be seen from Fig. 2(e) the splitting angle $\Delta\theta$ of $P_2$ decreases slightly as $K$ increases, indicating that the anisotropy has weak influence on the splitting angle in Landau structure.[4]

The dependence of magnetization distribution on exchange coefficient $A$ for the Landau structure is shown in Fig. 3, as $K$ = 500 J/m$^3$. The low proportion density in $V_{12}$ for the small $A$ (see Fig. 3(b)) means that a sharp domain wall with small net magnetization is formed between domain B and C, because the small $A$ allows large angle between neighbor magnetizations for forming the 90° domain wall. This is consistent with the variation of $P_1$ and $P_2$ shown in Fig. 3(a) and 3(c) respectively. The proportion densities of $P_1$, $P_{2L}$ ($P_{2R}$) and $V_2$ all increase with the decrease of $A$ as shown in Fig. 3(e). It can be seen from Fig. 3(d) that the splitting angle $\Delta\theta$ of $P_2$

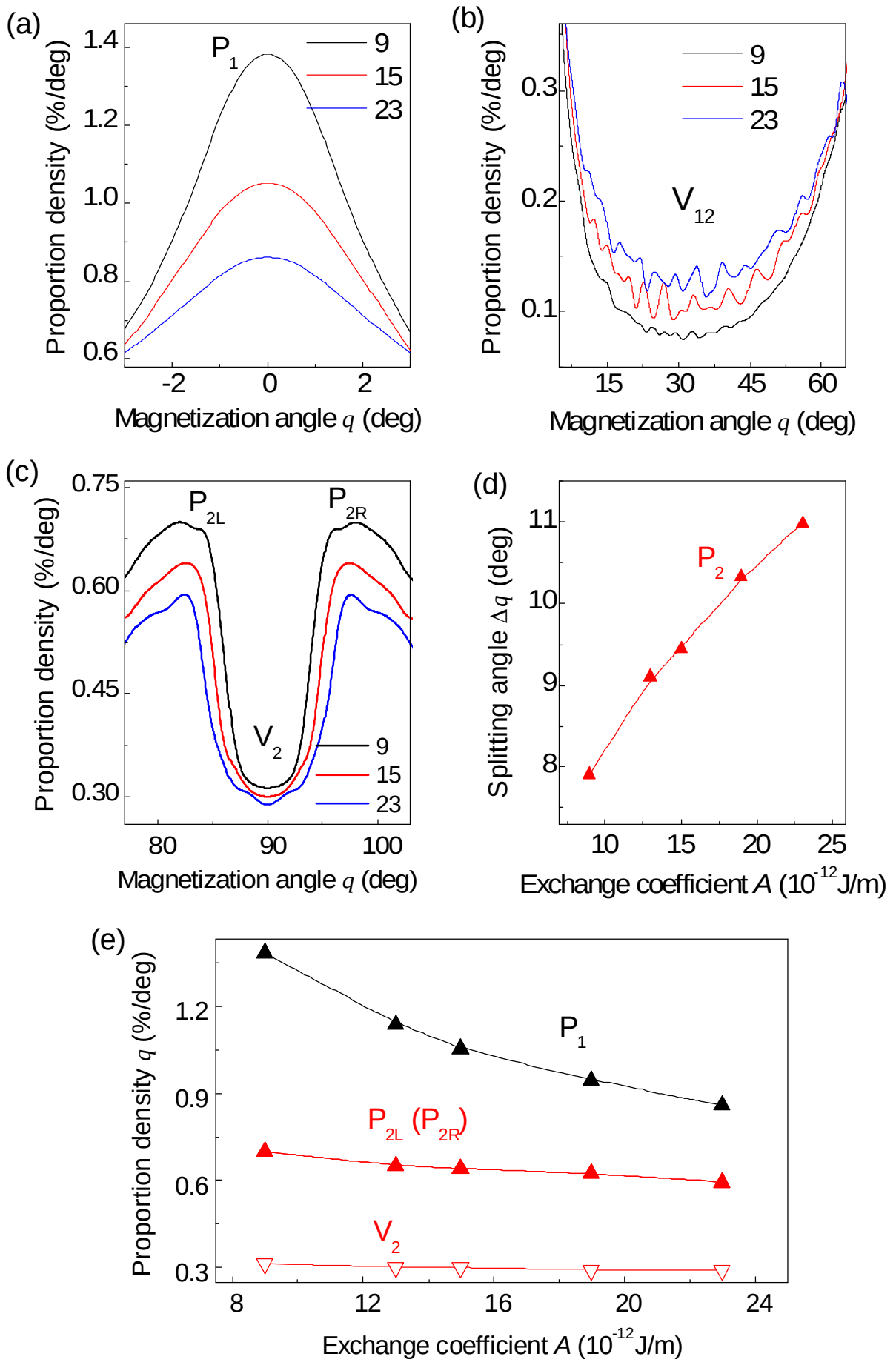

FIG. 3 (color online) For Landau domain structure, the distribution curves of of $P_1$ (a), $V_{12}$ (b), $P_2$ and $V_2$ (c) for $A = 9 \times 10^{12}$, $15 \times 10^{12}$ and $23 \times 10^{12}$ J/m, as $K = 500$ J/m$^3$. (d) Dependence of $\Delta\theta$ on $A$ for $P_2$. (e) Dependence of the proportion densities on $A$ for $P_1$, $P_{2L}(P_{2R})$, and $V_2$.

increases from 8° to 11° with $A$ from $9 \times 10^{12}$ J/m to $2.3 \times 10^{13}$ J/m, indicating much more influence than the anisotropy constant. It suggests the exchange interaction mainly contributes to the magnetization splitting in Landau structure.

## IV. DIAMOND DOMAIN STRUCTURE

Next, we show the results in diamond domain structure. Figure 4(a) shows a typical distribution curve of magnetization orientation as a function of the mangetization angle $\theta$ the in diamond domain structure. The distribution curve exhibits four main peaks $P_1$, $P_2$, $P_3$ and $P_4$ in a period of 360°, where show the symmetry to the short axis in diamond structure (see Fig. 1(d)). It is interesting that the magnetizaton splitting can be found in $P_1$, $P_2$ and $P_4$. As shown in Fig. 4(b), $P_1$ splits to two symmetric peaks $P_{1L}$ and $P_{1R}$, corresponding to the magnetization orientation in domain E and H in Fig. 1(d). However, $P_2$ whose magnetization around the 90° axis, splits to two asymmetric peaks $P_{2L}$ and $P_{2R}$, shown in Fig. 4(c). This asymmetry of the splitting could be explained by the different neighbor domains. For example, the splitting of $P_2$ describes the magnetization orientation of domain G and J. They both have a same neighbor domain K that show no splitting and locate in the center of rectangle. The other neighbor of domain G is domain H, while that of domain J is domain E respectively. It is noticed that the size, shape and boundary conditions of domain E are different from domain K and thus, the magnetization distribution in domain can be different due to the competition of the exchange and dipolar interactions.

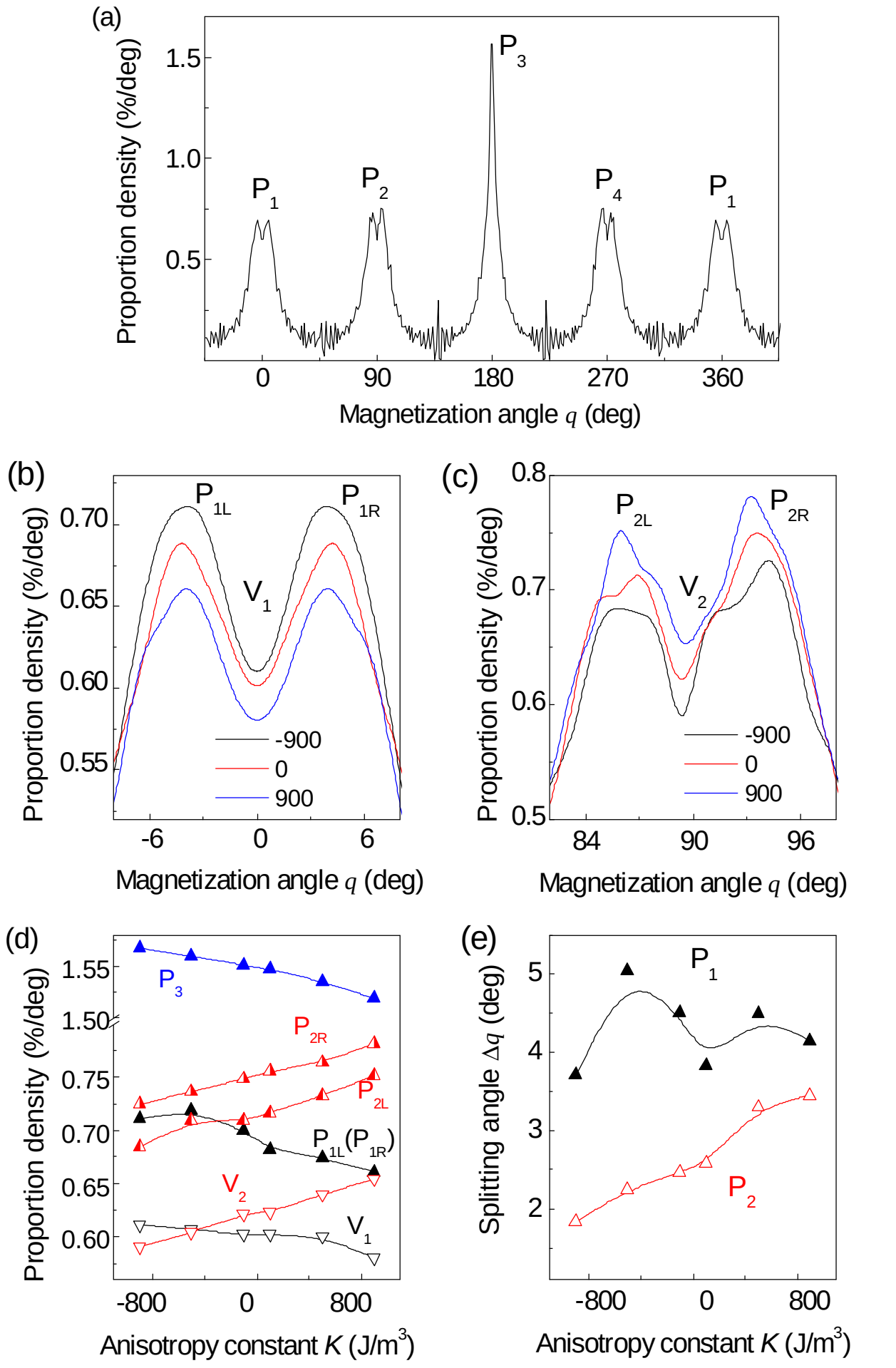

FIG. 4 (color online) For diamond domain structure: (a) the distribution curve of magnetization orientation as a function of the mangetization angle $\theta$. (b) the distribution curves of $P_1$ for $K = -900$, 0 and 900 J/m$^3$, as $A = 1.3 \times 10^{13}$ J/m. (c) the distribution curves of $P_2$ for $K = -900$, 0 and 900 J/m$^3$, as $A = 1.3 \times 10^{13}$ J/m. (d) Dependence of the proportion densities on $K$. (e) Dependence of the $\Delta\theta$ on $K$ for $P_1$ and $P_2$.

It can be seen from Fig. 4(d) that the proportion densities of $P_{1L}$ ($P_{1R}$), $V_1$ and $P_3$ decrease as the anisotropy constant $K$ increases. However, the proportion densities of $P_{2L}$, $P_{2R}$ and $V_2$ increase with $K$. This is due to the conservation of the total magnetization in the domain, i.e. wherever the magnetization of $P_1$ and $P_3$ gain, there must be partly a loss in that of $P_2$ and $P_4$. The magnetization favors aligning around the long axis as $K$ increases. The splitting angle $\Delta\theta$ of $P_1$ varies a little bit with $K$, as shown in Fig. 4(e), while that of $P_2$ increase with $K$.

The magentization distribution curves of $P_1$, $P_2$ and $P_3$ for various exchange coefficient $A$ are plotted in Fig. 5(a), (b) and (c). The proportion densities of $P_1$, $P_2$ and $P_3$

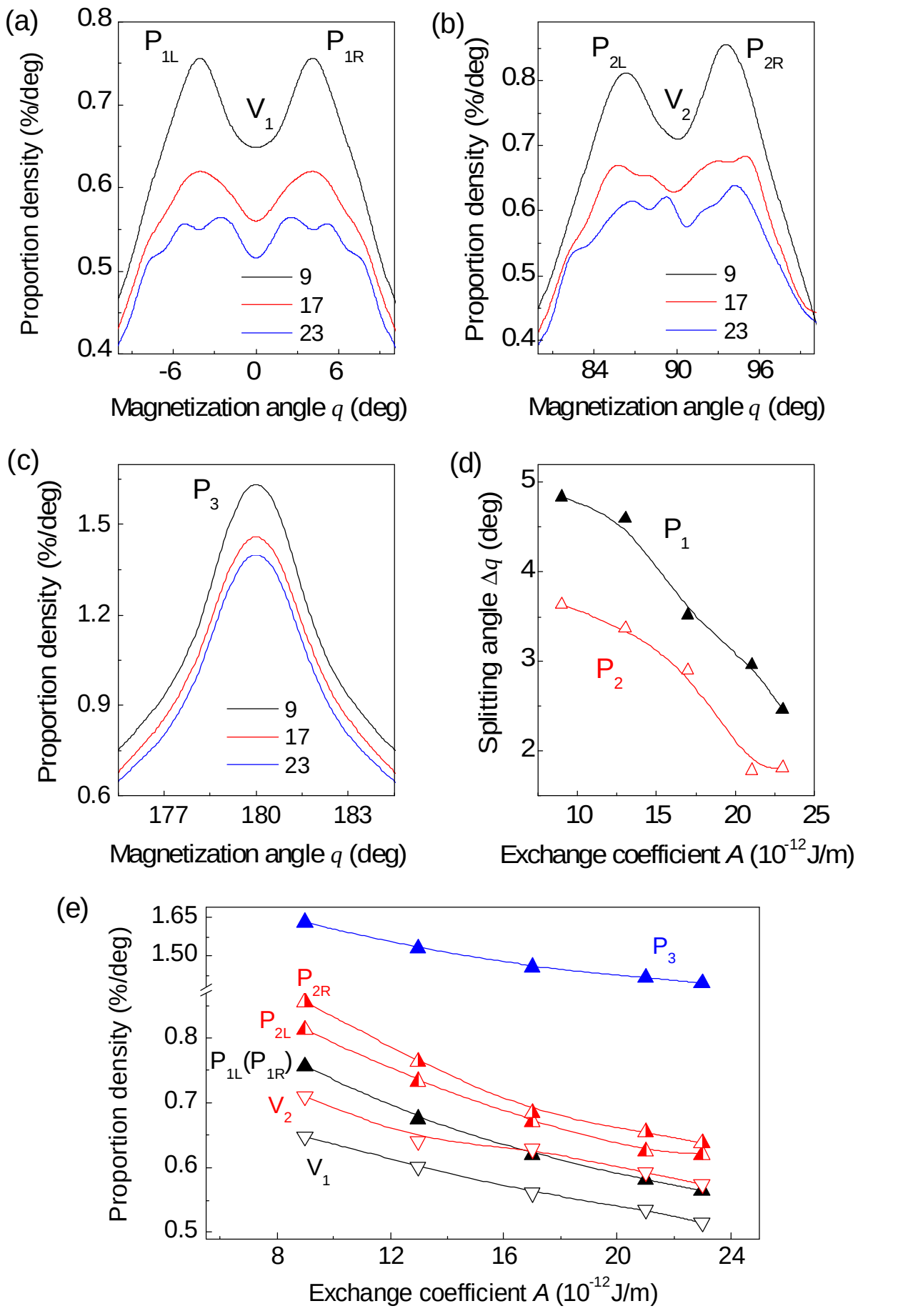

FIG. 5 (color online) For diamond domain structure, the distribution curves of of $P_1$ (a), $P_2$ (b) and $P_3$ (c) for $A = 9 \times 10^{12}$, $17 \times 10^{12}$ and $23 \times 10^{12}$ J/m, as $K = 500$ J/m$^3$. (d) Dependence of $\Delta\theta$ on $A$ for $P_1$ and $P_2$. (e) Dependence of the proportion densities on $A$.

decrease as increasing *A*, as shown in Fig. 5(e). As the exchange coefficient increases, sharp direction changes of neighbor magnetic moment are not allowed in the 90° domain wall region. This consumes the domains at both sides of the domain wall and makes the domains less magnetic moment. It can be seen from Fig. 5(d), $\Delta\theta$ for both $P_1$ and $P_2$ decrease with increasing *A*. This may be due to the change of volume charges created by the 180° Néel wall along the center,[4] and the decrease of the net magnetization in every domain as *A* increases. Due to the splitting angle decreases with increasing the exchange coefficient but increases with the anisotropy constant, we suggest that the magnetization splitting in diamond structure is resulted from magnetic anisotropy.

## V. SUMMARY

In summary, the distributions of magnetization orientation for both Landau and diamond domain structures in nano-rectangles have been investigated by micromagnetic simulation with various exchange coefficient and anisotropy constant. Both symmetric and asymmetric magnetization splitting are found in diamond domain structure, as well as only symmetric magnetization splitting in Landau structure. In the Landau structure, the splitting angle $\Delta\theta$ increases with the exchange coefficient *A* but decreases slightly with the anisotropy constant *K*, suggesting that the exchange interaction mainly contributes to the magnetization splitting in Landau structure. However in the diamond structure, the splitting angle $\Delta\theta$ increases with the anisotropy constant *K* but derceases with the exchange coefficient *A*, indicating that the magnetization splitting in diamond structure is resulted from magnetic anisotropy.

These results can extend the understanding of the magnetic domain microstructures. We expect the magnetization splitting in diamond domain structure could be observed in the future experiment.

This work was partly supported by CSKPOFR Grant No. 2009CB929503 and NSFC Grant No. 20833002.

*weiwei.lin@u-psud.fr

†haisang@nju.edu.cn

1. A. Lyberatos, G. Ju, R. J. M. van de Veerdonk, and D. Weller, J. Appl. Phys. **91**, 2236 (2002).
2. A. Krasyuk, F. Wegelin, S. A. Nepijko, H. J. Elmers, and G. Schönhense, Phys. Rev. Lett. **95**, 207201 (2005).
3. J. P. Park, P. Eames, D. M. Engebretson, J. Berezovsky, and P. A. Crowell, Phys. Rev. B **67**, 020403(R) (2003).
4. S. Hankemeier, R. Frömter, N. Mikuszeit, D. Stickler, H. Stillrich, S. Pütter, E. Y. Vedmedenko, and H. P. Oepen, Phys. Rev. Lett. **103**, 147204 (2009).
5. R. Hertel, O. Fruchart, S. Cherifi, P.-O. Jubert, S. Heun, A. Locatelli, and J. Kirschner, Phys. Rev. B **72**, 214409 (2005).
6. H. Kronmüller and R. Hertel, J. Magn. Magn. Mater. 215–**216**, 11 (2000).
7. M. Bode, A. Wachowiak, J. Wiebe, A. Kubetzka, M. Morgenstern, and R. Wiesendanger, Appl. Phys. Lett. **84**, 948 (2004).
8. A. Kobs, H. Spahr, D. Stickler, S. Hankemeier, R. Frömter, and H. P. Oepen, Phys. Rev. B. **80**, 134415 (2009)
9. K. Yu. Guslienko, W. Scholz, R. W. Chantrell, and V. Novosad, Phys. Rev. B **71**, 144407 (2005).
10. Y. Nakatani, A. Thivillé and J. Miltat, Nature Mater. **2**, 521 (2003).
11. J. K. Ha, R. Hertel, and J. Kirschner, Phys. Rev. B. **67**, 224432 (2003).
12. W. Rave and A. Hubert, IEEE Trans. Magn. **36**, 3886 (2000).
13. A. Krasyuk, S. A. Nepijko, A. Oelsner, C. M. Schneider and H. J. Elmers and G. Schönhense, Appl. Phys. A. **88**, 793 (2007).
14. S. Cherifi, R. Hertel, J. Kirschner, H. Wang, R. Belkhou, A. Locatelli, S. Heun, A. Pavlovska and E. Bauer, J. Appl. Phys. **98**, 043901 (2005).
15. D. Goll, G. Schütz, and H. Kronmüller, Phys. Rev. B. **67**, 094414 (2003).
16. A. Vansteenkiste, M. Weigand, M. Curcic, H. Stoll, G. Schütz and B. Van Waeyenberge, New J. Phys. **11**, 063006 (2009).
17. H. Min, R. D. McMichael, J. Miltat, and M. D. Stiles, Phys. Rev. B. **83**, 064411 (2011).
18. A. Hubert and R. Schäfer, *Magnetic domains: The analysis of magnetic microstructures* (Springer-Verlag, Berlin, 1998).
19. NIST Micromagnetic Modeling Activity Group.
20. S. Hankemeier, A. Kobs, R. Frömter, and H. P. Oepen, Phys. Rev. B. **82**, 064414 (2010).
21. M. J. Donahue and D. G. Porter, "OOMMF User's Guide, Version 1.0," NISTIR 6376, National Institute of Standards and

Technology, Gaithersburg, MD (Sept 1999), http://math.nist.gov/oommf/.

[22] W. H. Press, B. P. Flannery, S. A. Teukolsky, and W. T. Vetterling, *Numerical Recipies: The art of scientific computing* (Cambridge University Press, Cambridge, UK, 1986).